\documentclass[12pt]{article}

\usepackage{scicite}
\usepackage{graphicx}

\usepackage{times}

\newcommand{\apj}{{\it Astrophys. J.}}
\newcommand{\apjl}{{\it Astrophys. J.}}
\newcommand{\apjs}{{\it Astrophys. J. Suppl.}}
\newcommand{\aj}{{\it Astron. J.}}
\newcommand{\mnras}{{\it Mon. Not. R. Astron. Soc.}}
\newcommand{\aap}{{\it Astron. Astrophys.}}

\newcommand{\dmo}{dark-matter-only} 

\newenvironment{sciabstract}{%
\begin{quote} \bf}
{\end{quote}}

\newcounter{lastnote}
\newenvironment{scilastnote}{%
\setcounter{lastnote}{\value{enumiv}}%
\addtocounter{lastnote}{+1}%
\begin{list}%
{\arabic{lastnote}.}
{\setlength{\leftmargin}{.22in}}
{\setlength{\labelsep}{.5em}}}
{\end{list}}

\title{Stellar Feedback in Dwarf Galaxy Formation}

\author
{Sergey Mashchenko,$^{1\ast}$ James Wadsley,$^{1}$ H. M. P. Couchman$^{1}$\\
\\
\normalsize{$^{1}$Department of Physics and Astronomy, McMaster University,}\\
\normalsize{Hamilton, ON, L8S 4M1, Canada}\\
\\
\normalsize{$^\ast$To whom correspondence should be addressed; E-mail:  syam@physics.mcmaster.ca.}
}

\date{}

\begin{document} 




\maketitle


\begin{sciabstract}
Dwarf galaxies pose significant challenges for cosmological models. In
particular, current models predict a dark matter density that is
divergent at the center, in sharp contrast with observations which
indicate an approximately constant central density core. Energy
feedback, from supernova explosions and stellar winds, has been
proposed as a major factor shaping the evolution of dwarf galaxies. We
present detailed cosmological simulations with sufficient resolution
both to model the relevant physical processes and to directly assess
the impact of stellar feedback on observable properties of dwarf
galaxies. We show that feedback drives large-scale, bulk motion of the
interstellar gas resulting in significant gravitational potential
fluctuations and a consequent reduction in the central matter density,
bringing the theoretical predictions in agreement with observations.
\end{sciabstract}

Dwarf galaxies are the most common galaxy type\cite{mat98}.  In the
hierarchical picture of cosmic structure formation, dwarf galaxies
form first, later becoming building blocks for larger galaxies.
Thanks to their proximity in the local universe (around 18 galaxies
are located within 300~kpc of the Sun), several of these galaxies have
been studied in detail.  By measuring line-of-sight velocities for
hundreds of stars, accurate modeling of the mass distribution has
revealed features that pose severe challenges for the standard
cosmological model.  It appears, for example, that the distribution of
dark matter (which is the dominant mass component of these galaxies)
is of almost constant density in a central region that is comparable
in size to the stellar body of the
galaxy\cite{deb01,kle03,gen05,goe06,gil07}. In the best studied
systems, Fornax and Ursa Minor, the radius of this region is $\sim
400$~pc and $\sim 300$~pc, respectively\cite{gil07}.  This core is at
odds with existing cosmological models, which reliably predict the
dark matter to have a divergent density (a cusp) at the galactic
center\cite{nav04}.  Some dwarf spheroidal galaxies also exhibit
radial gradients in the stellar population, with stars more deficient
in heavy elements (and therefore presumed older) having a more
extended distribution and being kinematically warmer than more
metal-rich stars\cite{tol04,bat06}. Further, the presence of globular
clusters in many dwarfs is puzzling since these massive, compact
systems of many thousands of stars would have suffered gravitational
drag as they moved through the dark matter background of the galaxy
halo. This dynamical friction would have caused the globular clusters
to spiral into the galaxy center on time scales much shorter than the
age of the galaxy (the Fornax dwarf spheroidal galaxy is notable in
this regard\cite{goe06}).

It is well established that massive stars inject large amounts of
energy into the surrounding medium via stellar winds and supernova
explosions, resulting in large-scale (hundreds of parsecs in large
galaxies) random bulk motions of the interstellar gas at close to
sonic speeds ($\sim 10$~km~s$^{-1}$ for the typical gas temperature of
$10^4$~K)\cite{pel04,sly05,dea05}.  The effect of such perturbations
is larger for dwarf galaxies, as they have lower gas pressure due to
the lesser depth of their gravitational potential wells. This stellar
feedback has been invoked to explain at least some of the puzzling
properties of dwarf galaxies. In particular, there has been
considerable debate as to whether or not it can turn the theoretically
predicted central dark matter cusp into a
core\cite{nav96,gne02,rea05,mas06}.

Previous theoretical work has included both non-cosmological and
cosmological modeling.  High resolution, non-cosmological numerical
models with detailed descriptions of relevant physical
processes\cite{pel04,pel06,mar06} suffer from unrealistically
symmetric initial conditions and a static description of the dark
matter potential, and from the lack of gas accretion from the ambient
cosmic medium.  Prior attempts at self-consistent hydrodynamic
cosmological simulations have tended to focus on the formation of the
very first small galaxy progenitors\cite{abe02,yos03}, or on dwarf
galaxy models without sufficient resolution or the relevant physics to
properly model star formation and feedback because of the substantial
computational challenges involved in self-consistent
modelling\cite{ric05,rea06}.

Here we present the results of cosmological simulations of dwarf
galaxy formation and evolution that adequately resolve and model the
processes of star formation and stellar feedback. In good agreement
with our previous semi-analytical results\cite{mas06}, our
self-consistent model demonstrates that in small galaxies random bulk
gas motions driven by stellar feedback play a critical role in
determining the structure of the galactic center. The key result is
the transformation of the central density profile from a cusp to a
large core. This is a consequence of resonant heating of dark matter
in the fluctuating potential that results from the bulk gas
motions. We also demonstrate that the same mechanism can explain other
puzzling features of dwarf galaxies, such as the stellar population
gradients, low decay rate for globular cluster orbits and the low
central stellar density.

The simulations were run using the cosmological parallel tree code
Gasoline\cite{wad04}. This code represents dark and stellar matter as
a collection of dark matter and star particles and uses the smoothed
particle hydrodynamics (SPH) formalism to describe gas evolution.  A
detailed description of the code, including the prescriptions for star
formation and supernova feedback, can be found in
refs.\cite{sti06,suppl}. The very high resolution achieved in our
models required the addition of two key features to the standard
cosmological code. First, low temperature ($<10^4$~K) radiative
cooling from de-excitation of fine structure and metastable lines of
heavy elements was necessary to correctly model gas cooling in small
galaxies\cite{suppl}. Second, since our mass resolution
($<200$~M$_\odot$) is sufficient to resolve individual supernovae, we
introduced a new, stochastic, prescription for stellar
feedback\cite{suppl}.

We created cosmological initial conditions with input constraints
designed to produce a dwarf galaxy with total mass $\sim
10^9$~M$_\odot$ at redshift $z=6$ within a box of size 4 co-moving
Mpc.  A central, high resolution, sphere with radius 0.4 co-moving Mpc
was populated with gas particles. The particle masses inside the high
resolution sphere were 1,900~M$_\odot$ for dark matter and
370~M$_\odot$ for gas. The mass of particles generated to represent
stars was $\sim 120$~M$_\odot$. At the end of the simulations, the
total numbers of dark matter, gas, and star particles were $1.1\times
10^7$, $4.5\times 10^6$, and $4.5\times 10^5$, respectively.  The
gravitational softening length was held constant at 12~pc.  Further
model details, including the description of numerical convergence
tests and free parameter studies, can be found in
\cite{suppl}.

Two primary simulations were run. The first one included all the key
physical effects: gas dynamics, star formation, and stellar
feedback. (This was by far the most computationally expensive,
consuming $6\times 10^5$ CPU hours.)  The second one was a \dmo\
control simulation. The simulations started at $z=150$, and ended at
$z=5$.

\begin{figure}
\begin{center}\includegraphics[scale=0.35]{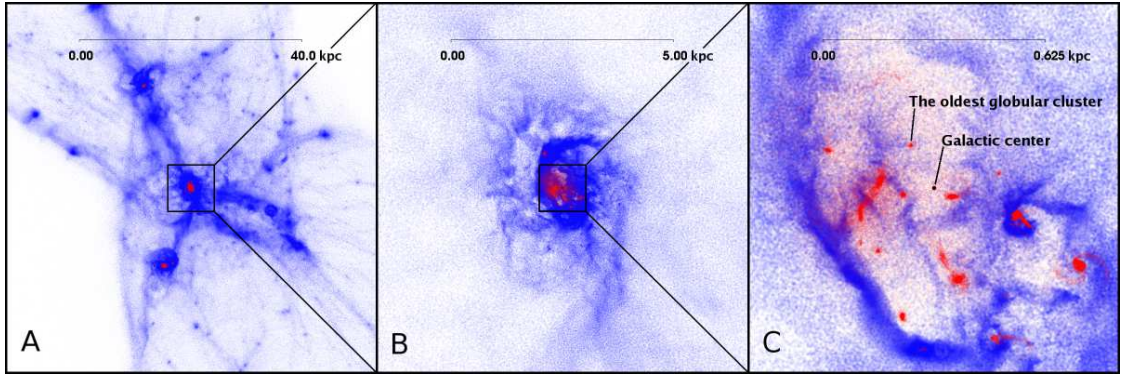}\end{center}
\caption{A zoom-in on the central part of the simulated forming dwarf galaxy
at redshift $z=5.3$. This time was chosen to illustrate the very
clumpy gas distribution following a star burst. Gas is shown in blue
and stars in red. ({\bf A}) Global view.  ({\bf B}) View of the
galaxy. ({\bf C}) The central part of the galaxy.  A number of star
clusters are visible in panel {\bf C}, the oldest (marked) has an
age of 200~Myr.}
\end{figure}

In the simulations, the matter distribution develops the classic
web-like or filamentary structure on large scales, with the most
massive galaxy forming around $z=10$ at the intersection of the major
filaments near the center of the computational box (Figure~1). The
evolution of the galaxy is relatively smooth (no major mergers)
between $z=8$ and 5. The star formation in the dwarf galaxy is very
bursty, with major star bursts repeating approximately every $\sim
80$~Myr, consistent with the non-cosmological models of \cite{pel04}.
Stars form predominantly in clusters, but many of them quickly
disperse. Starting at $z=6.2$, when the galactic stellar mass reaches
$\sim 10^7$~M$_\odot$, clusters which survive until the end of the
simulation start forming. These long-lived clusters have broadly the
same sizes ($\sim 10$~pc, essentially unresolved in our simulations),
masses ($\sim 10^5$~M$_\odot$), and heavy element abundance ($\sim
3$\% of solar) as globular clusters observed in the local universe.
It is noteworthy that in the Local Group no old (early-type) dwarfs
with stellar mass $<10^7$~M$_\odot$ have globular clusters, whereas
all brighter dwarfs (with the exception of M32, which is severely
tidally stripped by its host galaxy, M31) have globular
clusters\cite{mat98}. This suggests that a galaxy has to be large
enough ($> 10^7$~M$_\odot$ in baryons) to produce globular clusters,
in good agreement with our simulations.

\begin{figure}
\begin{center}\includegraphics[scale=0.5]{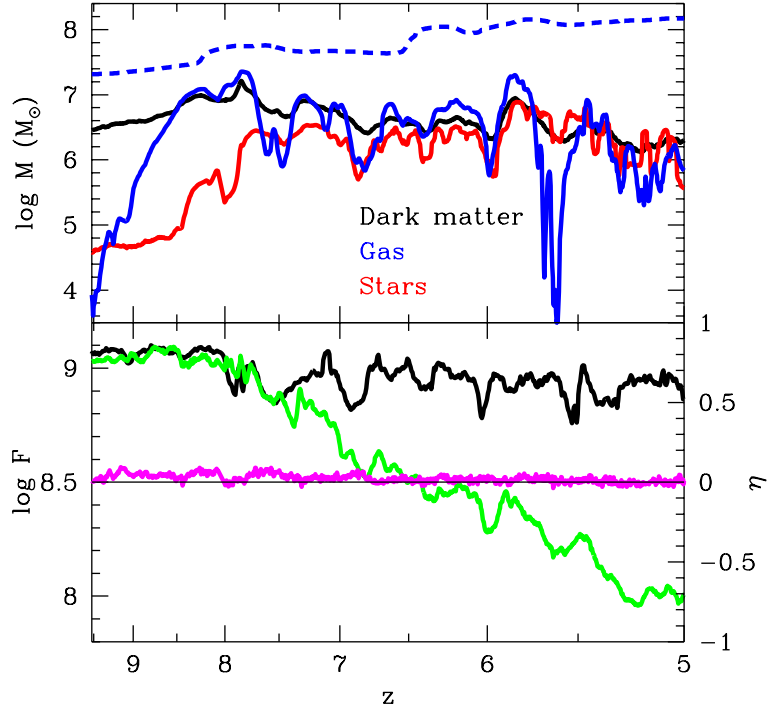}\end{center}
\caption{Evolution of the central quantities in the model dwarf galaxy.
In the upper panel, solid lines correspond to changes in the dark
matter (black), gas (blue), and stellar (red) masses enclosed within
the central 100~pc as a function of the redshift, $z$. The dashed blue
line shows the evolution of the enclosed gas mass within the central
1.6~kpc (half the virial radius). In the lower panel, green and black
lines show the evolution of the central dark matter phase-space
density, $F$, for the hydrodynamic and \dmo\ simulations,
respectively. We also show the evolution of the velocity anisotropy,
$\eta$, for the same dark matter particles as were used to calculate
$F$ (magenta line; horizontal black line marks $\eta=0$).  Here
$\eta\equiv(\sigma_r^2-\sigma_t^2)/(\sigma_r^2+\sigma_t^2)$, where
$\sigma_r$ and $\sigma_t$ are, respectively, the one-dimensional
radial and tangential velocity dispersions.}
\end{figure}

Feedback from the bursty and clustered star formation results in a
dramatically perturbed interstellar gas distribution on large scales
(hundreds of parsecs; Figures~1 and 2 and Movie~S1). This is
consistent with the observed (irregular) distribution of gas in dwarf
galaxies\cite{mat98}. Note that at high-redshift this feedback does
not expel gas from the galaxy, in contrast to the maximum stellar
feedback mechanism\cite{gne02}. Instead, supernova explosions compress
gas into large shells and filaments, which are confined to the central
part of the galaxy and move with speeds $\sim 10-20$~km~s$^{-1}$
(comparable to the speeds of dark matter particles). We showed
previously\cite{mas06} that gas motion with these characteristics
results in efficient gravitational heating of the central dark matter
and flattening of the cusp.

\begin{figure}
\begin{center}\includegraphics[scale=0.5]{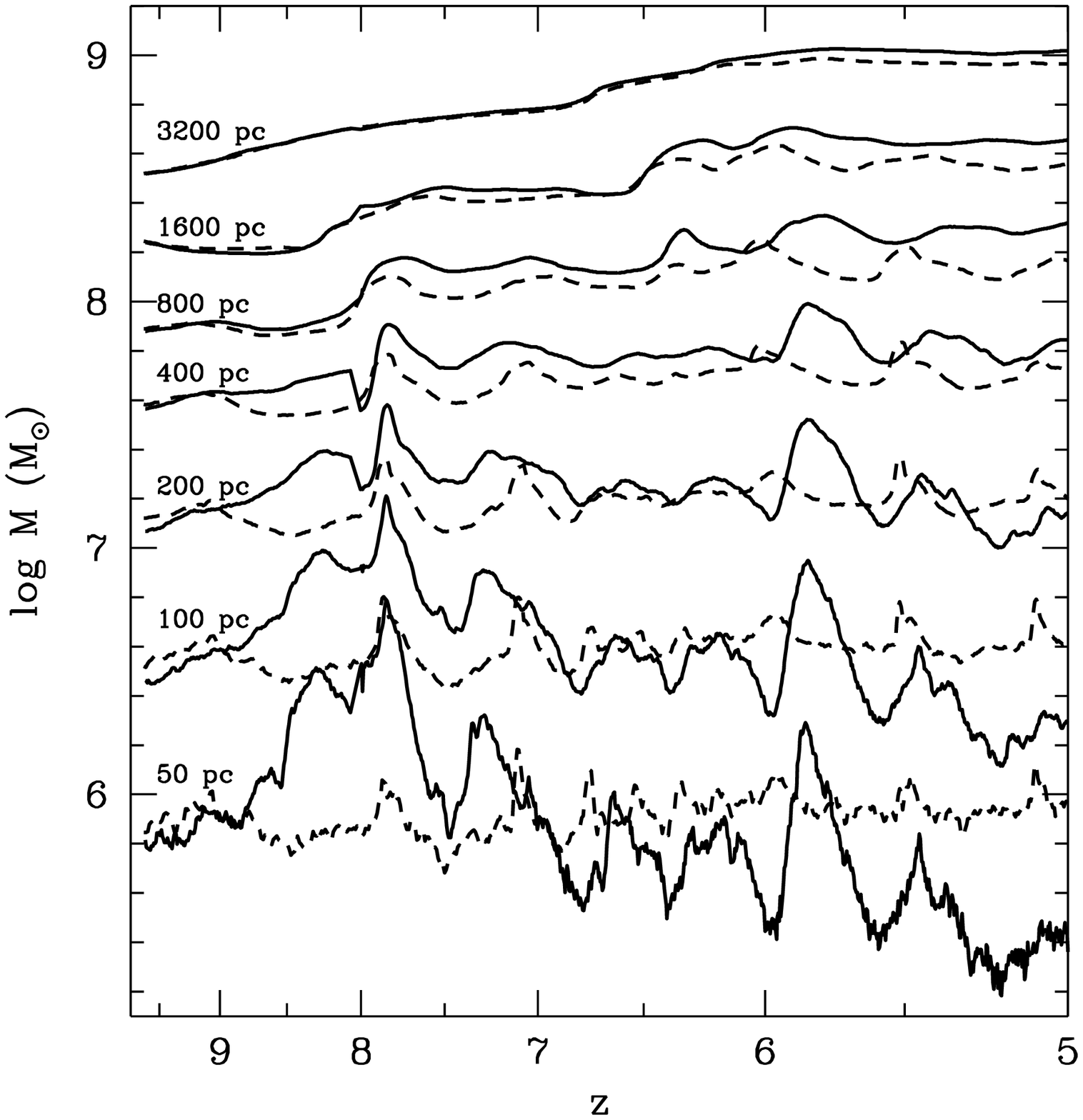}\end{center}
\caption{Evolution of the enclosed dark matter masses in the model 
galaxy at different radii. Dashed lines correspond to the \dmo\
simulation, and solid lines correspond to the hydrodynamic
simulation.}
\end{figure}

\begin{figure}[t]
\begin{center}\includegraphics[scale=0.5]{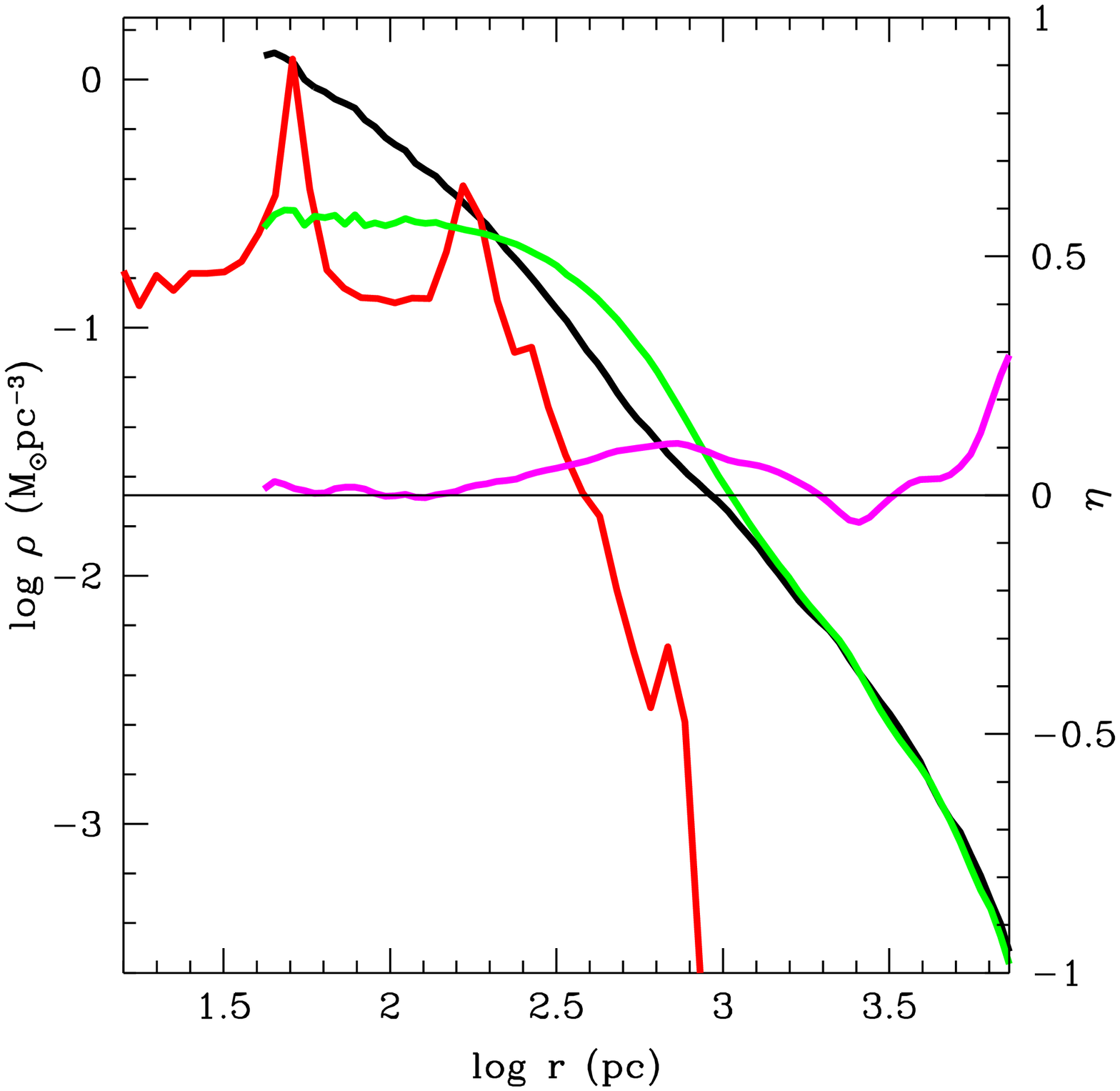}\end{center}
\caption{Radial profiles for the model galaxy at redshift $z=5.2$. 
At this time the central gas density is very low, minimizing the
adiabatic compression of dark matter due to baryons (which makes it
appropriate for comparison with presently observed gas-poor dwarfs).
Green and red lines show the dark matter and stellar density ($\rho$)
profiles, respectively, in the hydrodynamic simulation. The thick
black line corresponds to the dark matter density profile for the
\dmo\ simulation. The magenta line shows the velocity anisotropy,
$\eta$ (see caption to Figure~2 for the definition), profile for the
dark matter (in the hydrodynamic simulation).}
\end{figure}

The heating of dark matter in our model dwarf galaxy is highly
effective (Figures 2--4). Whereas both density, $\rho$, and velocity
dispersion, $\sigma$, of the particles are strongly affected by the
variable content of gas and stars at the galactic center, the
phase-space density, $F=\rho/\sigma^3$, is much less sensitive to
adiabatic compression of dark matter by baryons (Figure~2). In the
\dmo\ simulation, $F$ stays approximately constant, whereas in the
hydrodynamic simulation, $F$ gradually decreases with time as the
result of the stellar feedback, becoming a factor of 10 lower than for
the \dmo\ simulations at the end of the evolution.

The dark matter density is strongly affected by the stellar feedback
only in the central region of the galaxy (Figure~3). This is the
region in which the enclosed gas mass occasionally dominates that of
the dark matter, and is where the gas is most strongly affected by the
feedback. At the end of the hydrodynamic simulations, the dark matter
density at the smallest resolved radius becomes a factor of seven
smaller than in the \dmo\ simulations.

Whereas the dwarf galaxy halo in the \dmo\ simulation develops a
central cusp with logarithmic slope of $\gamma=-0.95$, consistent with
previous predictions of the standard model\cite{nav04}, in the
hydrodynamic simulations, resonant heating due to stellar feedback
turns the cusp into a flat core with radius $400$~pc (Figure~4) and
average density $0.2$~M$_\odot$~pc$^{-3}$. These core parameters are
close to those inferred for Fornax, $\sim 400$~pc\cite{gil07} and
$\sim 0.1$~M$_\odot$~pc$^{-3}$\cite{mat98}, respectively.  The same
mechanism produces a core of somewhat smaller radius ($\sim 300$~pc)
in the distribution of stars, and, significantly, pushes newly formed
globular clusters away from the galactic center. The four oldest
globular clusters, for example, were born with radial distance
dispersion $\sigma_r=37$~pc (essentially at the galactic center), but
after $\sim 200$~Myr of evolution this distance had grown to a
time-averaged value of $\sigma_r=280$~pc (comparable to the stellar
core radius). We suggest that resonant gravitational heating can at
least partially explain why globular clusters in Fornax, and in some
other dwarfs, are located at large distances from the galactic
center\cite{goe06}. Two mechanisms contribute to the effect: first,
the feedback flattens the central cusp, which reduces the efficiency
of dynamical friction in the central regions\cite{goe06}; second,
stellar feedback would have continued to heat the globular cluster
orbits until stars stopped forming, around 200~Myr ago in
Fornax\cite{bat06}.

The distribution of velocities is isotropic within the core, and shows
slight radial anisotropy outside the core (Figure~4), whereas the core
remains isotropic throughout the evolution (Figure~2). This behavior
is inconsistent with a mechanism\cite{elz01} employing massive gas
clouds, passively orbiting (not driven by feedback) near the galactic
center, which flatten the dark matter cusp via heating due to
dynamical friction. It has been shown\cite{ton06} that this would
result in the development of significant tangential anisotropy within
the core, which is not observed in our simulations. On the other hand,
the gravitational resonance heating\cite{mas06} naturally produces
isotropic cores due to the fact that the feedback-driven bulk gas
motions have random directions.

These results also provide a natural explanation for the stellar
population gradients seen in many early-type
dwarfs\cite{tol04,bat06}. In our simulations star formation is
concentrated toward the galactic center. Over time, feedback gradually
heats the population of stars, resulting in older (and more
metal-poor) stars being kinematically warmer and having a larger
spatial extent than younger (and more metal-rich) stellar
populations. Hence we can reproduce, qualitatively, the age,
metallicity, and velocity dispersion gradients observed in dwarf
galaxies.

Our simulations were stopped at $z=5$ as continuing beyond this point
would require a much larger computational box to correctly model the
growth of larger structures and an infeasible increase in computation
time. Furthermore, the impact of external ionizing radiation, ignored
in our model, can become significant after $z=6.5$. Nevertheless, we
can reasonably infer the subsequent evolution of our model galaxy. If
it is to become one of the early-type galaxies in the local universe
(which are gas-poor), some mechanism will have to remove most or all
of its interstellar medium. Some combination of a powerful star burst,
increased metagalactic ionizing radiation, and ram-pressure stripping
could result in the dwarf losing most of its gas\cite{may07}. It is
also likely that only a fraction of its star clusters will survive
until the present time.  As a result, our model galaxy would end up
resembling a large dwarf spheroidal galaxy in the local universe: low
stellar density; metal-poor with old stellar populations having
pronounced radial population gradients; large stellar and dark matter
cores (comparable in size and density to those in dwarf spheroidals);
and perhaps a few globular clusters. In many respects, the galaxy
would resemble the Fornax dwarf.

Our non-cosmological modeling\cite{mas06} suggested that stellar
feedback can be directly responsible for the absence of dark matter
cusps only in small galaxies, with total masses $<10^{10}$~M$_\odot$:
in larger galaxies the dark matter particle velocities become
significantly larger than the velocity of the random gas bulk motions,
$\sim 10$~km~s$^{-1}$. Our current, cosmological simulations are
consistent with this result (the mass of our galaxy reaches $2\times
10^9$~M$_\odot$ by $z=5$). Numerical simulations \cite{kaz06} have
suggested that a universal halo density profile (either cuspy or
cored), once set is preserved through subsequent hierarchical
evolution (which is consistent with the analytical result that the
collisionless dark matter phase-space density can only decrease over
time), implying that our mechanism may also lead to dark matter cores
in larger galaxies.

Our simulations indicate that the gravitational heating of matter
resulting from feedback-powered bulk gas motions is a critical
determinant of the properties of dwarf galaxies. Large dark matter
cores are an unavoidable consequence of early star formation in dwarf
galaxies. Our model indicates that, in primordial dwarf galaxies,
globular clusters are formed in the most natural place---near the
center, where the gas pressure is highest---and are then pushed by
feedback to much larger distances. This mechanism also ensures that
globular clusters and unclustered stars have a comparable
distribution, as is observed in early-type dwarfs\cite{lot01}.
Additionally, the low stellar density and stellar population gradients
observed in dwarf galaxies are also expected from the model.  Finally,
it is also worth noting that large cores have serious implications for
direct searches of dark matter, as a flat core will produce a much
weaker gamma ray annihilation signal than a cusp.

\begin{scilastnote}
\item The simulations reported in this paper were carried out on
  facilities of the Shared Hierarchical Academic Research Computing Network
  (SHARCNET:www.sharcnet.ca). The authors would like to acknowledge the support
  of the Canadian Institute for Advanced Research, NSERC, and SHARCNET. To
  produce our Figure~1 and Movie~S1 we used the public domain program IFrIT
  written by Nick Gnedin (Fermilab).
\end{scilastnote}

\medskip

{\bf Supporting Online Material\\}
Methods\\
Figs. S1, S2, S3, S4, S5\\
References\\
Movie S1


\clearpage
\renewcommand{\thefigure}{S\arabic{figure}}
\setcounter{figure}{0}

\begin{center}
\LARGE \bf
Supporting Online Material\\

\bigskip

Methods
\end{center}

\section*{Model}

We used GASOLINE\cite{wad04s}, a parallel TreeSPH particle code that
has individual particle timesteps.  In several respects our run
parameters were similar to those used for the galaxy formation
simulations run with GASOLINE reported in \cite{gov07}.  In what
follows we will highlight the differences required to simulate small
galaxies at high redshift. The code includes Compton cooling, atomic
cooling based on collisional equilibrium and low-temperature cooling
due to metals.  The cosmic ultraviolet background is assumed to be low
at the redshifts we considered\cite{haa96} and was omitted.

We model the low-temperature ($<10^4$~K) radiative cooling of gas
through fine structure and metastable lines of C, N, O, Fe, S, and Si
following the prescription in \cite{bro01s}. To derive the cooling
function, the authors assumed that the above elements are maintained
in ionization equilibrium by locally produced cosmic rays, and that
the cosmic ray ionization rate is scaled from the Galactic value by
$Z/Z_\odot$, where $Z/Z_\odot$ is the metallicity of gas in solar
units.  We found that for temperatures $T=20\dots 10^4$~K their
cooling function can be approximated very well by the following
empirical expression:

$$
\log (\Lambda/n_{\rm H}^2) = -24.81 + 2.928x - 0.6982x^2 + \log(Z/Z_\odot),
$$

\noindent where $x\equiv \log(\log(\log(T)))$. Here
$n_{\rm H}$ is the number density of hydrogen atoms (in cm$^{-3}$),
and $\Lambda/n_{\rm H}^2$ is in erg~cm$^3$~s$^{-1}$ units.

The star formation algorithms we used were very similar to those that
were extensively tested by Stinson et al.\cite{sti06s} and Governato
et al.\cite{gov07}.  In order for gas to form stars it has to be Jeans
unstable, to be colder than $15,000$ K, to have a minimum physical
density of $10$ atoms/cc and to have a minimum overdensity of 1,000.
Star creation would then proceed at an average rate, in terms of solar
masses per unit volume per unit time, of 0.05 times the gas density
divided by the dynamical time.  Star particles were created
stochastically, as described in \cite{sti06s}, with a mass of 120
solar masses each.  The probability that a star would be created in
any million year interval was proportional to the average star
formation rate of the parent gas particle multiplied by the time
interval divided by the star mass.

A Kroupa\cite{kro98} initial stellar mass function was used to
determine the rate of supernovae per solar mass of stars.  Each
supernova was assumed to deposit $E_{\rm SN}=0.4\times10^{51}$~ergs
into the gas in net (allowing for some loss in the process).  Our high
mass resolution made it inappropriate to average the feedback effects
due to a large population of stars when describing the feedback due to
an individual star particle.  Instead, we employed stochastic feedback
by translating the time-averaged feedback into a probability that an
individual supernova would occur in that time.  Thus the energy is
injected in discrete supernova events.  For larger star particle
masses the energy output converges to the result with continuous
feedback.  Star particles select nearby gas particles in the manner
described in \cite{sti06s} to receive feedback energy.  However, the
energy was added in a volume-weighted rather than mass-weighted manner
which better respects the isotropic nature of the energy injection.

\section*{Initial conditions}

In our simulations, we used the following values of the cosmological
parameters\cite{spe03}: matter density $\Omega_m=0.27$, baryonic
density $\Omega_b=0.044$, Hubble constant
$H_0=71$~km~s$^{-1}$~Mpc$^{-1}$, amplitude of fluctuations
$\sigma_8=0.84$, and spectral index $n_s=0.93$. We assumed that the
universe is flat ($\Omega_m+\Omega_\Lambda=1$, where $\Omega_\Lambda$
is the cosmological constant.)

Constrained cosmological initial conditions were constructed using the
package COSMICS\cite{ber95}. We used two concentric spherical Gaussian
ball constraints.  The first constraint had a Gaussian scale length of
$R_1=0.119$ co-moving Mpc and an initial overdensity of
$\delta_1=1.686 D(0)/D(z_{\rm coll})$, where $D(z)$ is the linear
growth factor, and $z_{\rm coll}=5.92$ is the targeted collapse
redshift. This was designed to produce a halo with the mass of $\sim
10^9$~M$_\odot$ by $z\sim 6$. After a few tests we realized that for
our large box size (4 co-moving Mpc) one constraint was not enough to
produce an isolated dwarf galaxy by $z=5$. We chose a second
constraint, with scale length of $R_2=0.357$ co-moving Mpc and initial
overdensity of $\delta_2=\delta_1 [0.5 (R_2/R_1)^2 + 0.5]^{3/2}=0.0894
\delta_1$. (The value of $\delta_2$ assumes an isolated density peak
in an empty universe.)

We used COSMICS to produce a cube populated with $1024^3$ dark matter
particles at $z_0=150$. The central sphere, with radius of 0.4
co-moving Mpc and containing $\sim 4.6\times 10^6$ dark matter
particles, was then populated with the same number of gas particles
(each gas particle initially overlaying a dark particle and being
assigned the same velocity).  Following
\cite{ann96}, the initial temperature of gas was set to $2.73
(1+z_0)^2/(1+200)$~K$=310$~K. (The above expression assumes that the
temperature of gas is the same as that of cosmic microwave background
radiation until $z=200$ and is set by adiabatic expansion thereafter.)

To make the simulations computationally feasible, we lowered the
resolution outside the central sphere by averaging coordinates and
velocities for groups of adjacent particles. Between radii 0.4 and 0.8
co-moving Mpc, the number of particles was reduced by a factor of
$2^3$, and in the remainder of the computational box by a factor of
$8^3$. These low-resolution particles are needed to reproduce tidal
torques from large scales on structures inside the central
high-resolution sphere. Particles in the low resolution regions are
collisionless and account for the mass of both dark matter and gas in
those regions.

The standard numerical code does not have the required physics (such
as non-equilibrium molecular hydrogen chemistry and radiative
transfer) to self-consistently describe star formation in small
progenitor galaxies with virial temperature $T_{\rm
vir}<10^4$~K. Proper modeling of these galaxies is important as they
pre-enrich the intergalactic medium (out of which our model dwarf
galaxy will be built) with trace amounts of heavy elements.  We
circumvented this limitation by running the hydrodynamic simulations
with a simplified prescription for star formation until $z=9.6$
(before our model galaxy was assembled), and then switching to our
default star formation recipe.  The simplified portion of the
simulations had no low-temperature ($<10^4$~K) cooling due to heavy
elements, and employed simple gas density (overdensity $>100$ and
density $>1$~cm$^{-3}$) and temperature ($T<3\times 10^4$~K) criteria
to select star-forming particles. At $z=9.6$, our model galaxy had a
heavy element abundance of $\sim 0.001$ solar units, which is
comparable to the lower stellar metallicity cutoff observed in nearby
early-type dwarf galaxies\cite{hel06}. (At the end of the simulations,
the average metallicity of gas in the model galaxy reached 0.014 solar
units.) Despite these significant simplifications, in the initial
phase of the simulations we produce the correct abundance of heavy
elements in the intergalactic medium. We also explored a range of
heavy elements abundances at $z=9.6$, by scaling the metallicity of
all gas particles by a constant factor in the range $\xi=0.2$ to 5
(see the next section).  After $z=9.6$, we ran the simulations with
the full physics (low-temperature cooling and Jeans instability
criterion).

\section*{Numerical tests}

To estimate the smallest resolved radius at the center of our model
galaxy, we carried out additional \dmo\ simulations which had 8 times
lower mass resolution than our fiducial model. The initial conditions
for this run were generated from the initial conditions for the high
resolution run by binning together groups of 8 adjacent particles
(reducing the effective resolution by a factor of two in each linear
dimension). We chose the gravitational softening length for the new
run to be twice larger ($\varepsilon=24$~pc) than for the fiducial
run.

\begin{figure}[h]
\begin{center}\includegraphics[scale=0.5]{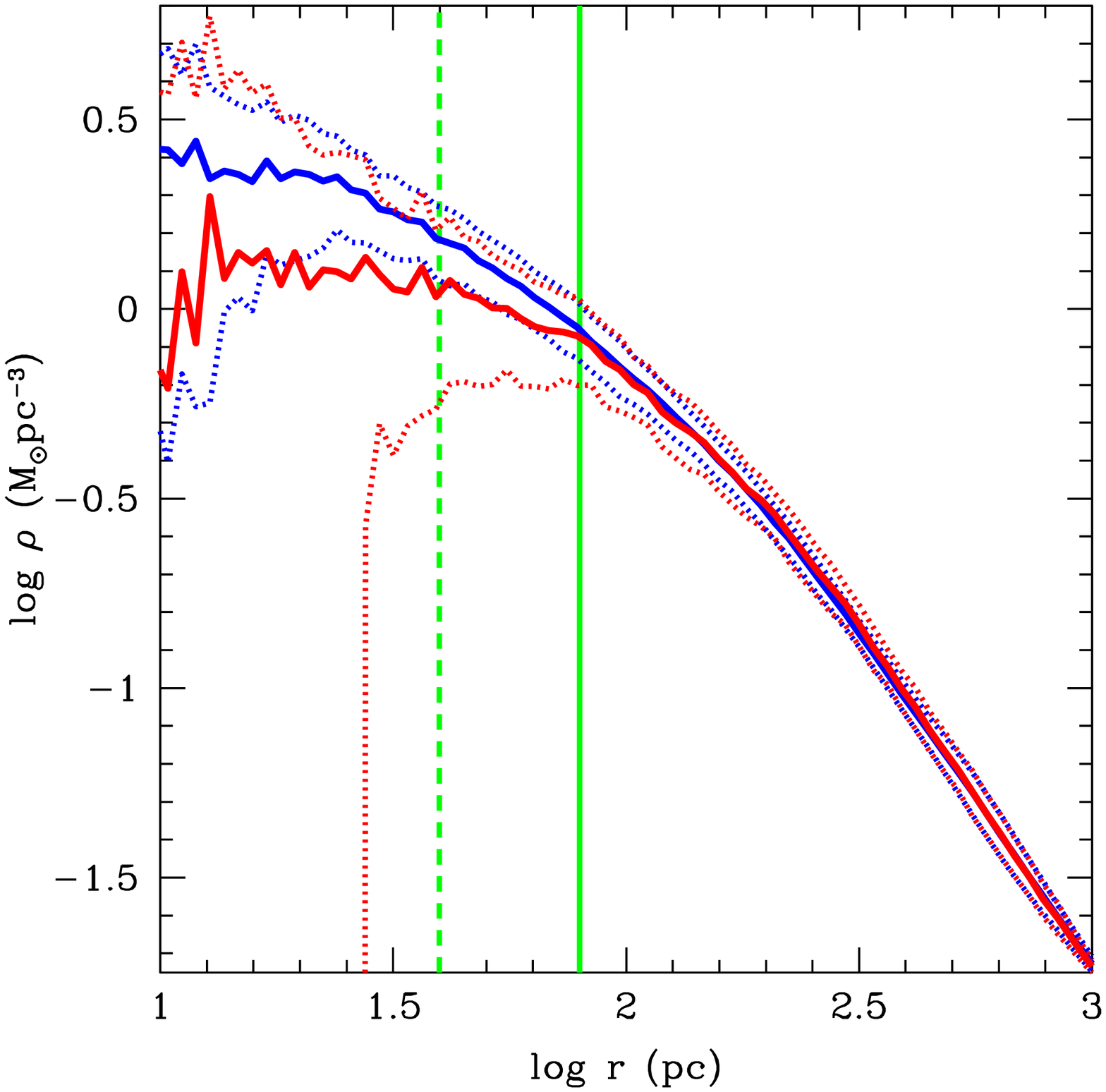}\end{center}
\caption{Radial density profiles for the model galaxy for \dmo\ simulations at $z=5$.
Solid blue and red lines show the dark matter density profiles
averaged over the last 50 snapshots for the fiducial model and the
model with 8 times lower mass resolution, respectively. The dotted
blue and red lines show the corresponding one-sigma deviations from
the averaged profiles. The vertical solid green line is at 80~pc, the
radius where the low and high resolution models converge. The vertical
dashed green line corresponds to the estimated smallest resolved
radius (40~pc) for the high resolution run.}
\end{figure}

Figure~S1 shows the final radial density profiles for both high and
low resolutions runs. One can see that both profiles converge around
radius $\sim 80$~pc (or $3.3 \varepsilon$), implying that our high
resolution model should resolve radii down to $\sim 40$~pc. The actual
resolution is probably somewhat smaller in the fiducial run. At the
radius where the low and high resolution models start diverging, the
low resolution run exhibits both a break in the density profile
(becomes shallower) and the one-sigma spread becomes significantly
wider (see Figure~S1). For the high resolution run, the same behavior
is observed at a radius of 25~pc, or $2
\varepsilon$. Thus we are confident that in the fiducial models we resolve
central radii to at least 40~pc, and perhaps to 25~pc. This is
sufficient for our purposes, as stellar feedback significantly
perturbs the distribution of galactic gas on scales $> 100$~pc.

Next we explored model sensitivity to the three important model
parameters: $E_{\rm SN}$ (the energy deposited into the interstellar
medium by a single supernova); the low-temperature cooling factor,
$\kappa$ (see the description below); and the initial metallicity
factor, $\xi$. (The fiducial values are $E_{\rm SN}=0.4\times
10^{51}$~ergs, $\kappa=1$, and $\xi=1$.) It would be computationally
prohibitive to run full-scale simulations for different values of the
above parameters.  Instead, we ran a sequence of small-box
simulations. At $z=9.6$, we cut out a sphere centered on our model
galaxy with radius 6.1~kpc (three virial radii at that epoch). We then
ran eight different cosmological simulations, with the extracted
sphere placed in an empty universe. Two of the runs had identical
parameters to the fiducial full-box simulations (a \dmo\ run and a
hydrodynamic run). Six other small-box simulations were run with two
different values for each of the three free parameters, with the rest
of the parameters kept at their fiducial values.

\begin{figure}
\begin{center}\includegraphics[scale=0.5]{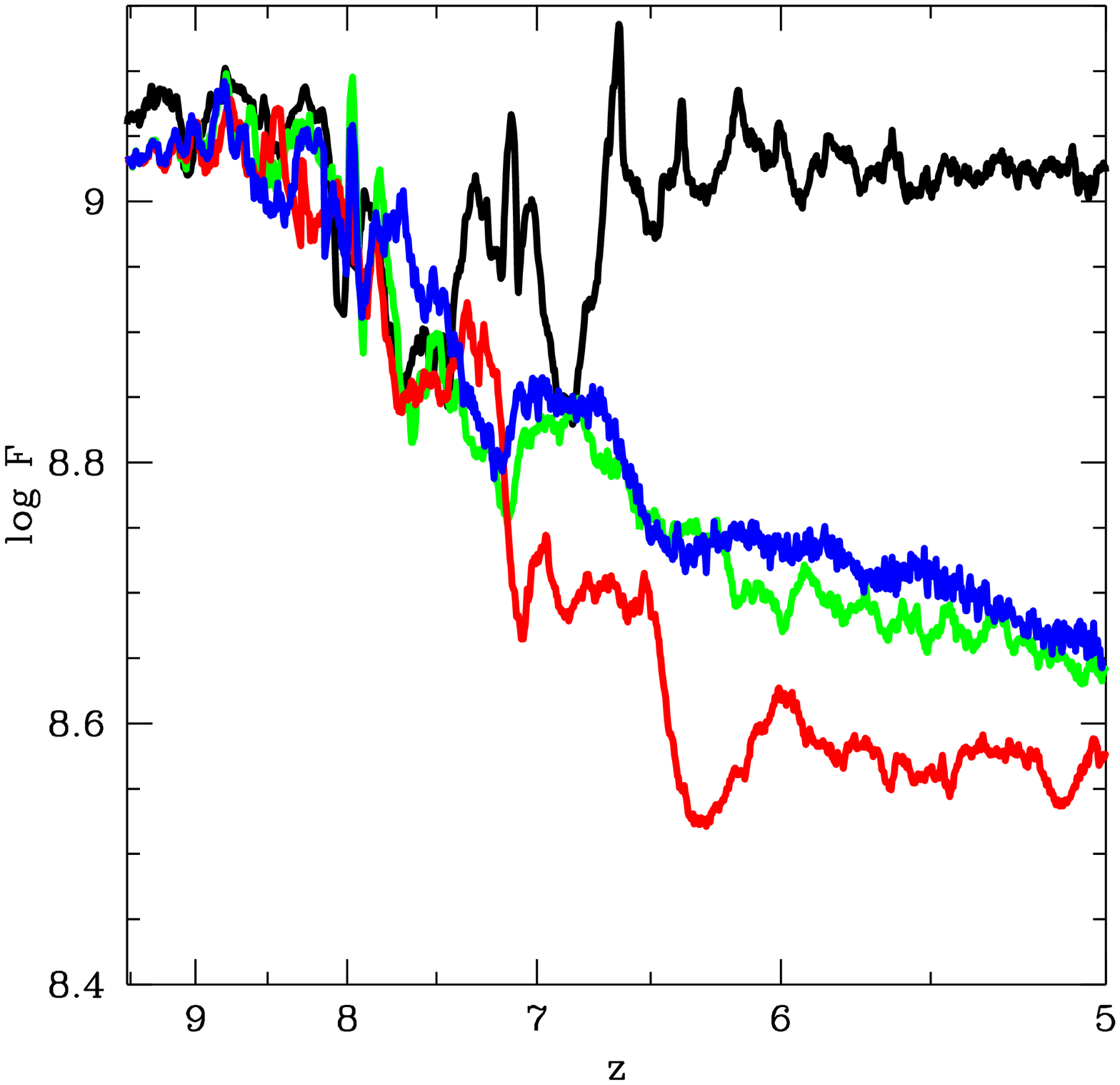}\end{center}
\caption{Evolution of the central dark matter phase-space density, $F$, for 
small-box simulations with different values of $E_{\rm SN}$. Black and
green lines correspond to the \dmo\ and hydrodynamic fiducial models
($E_{\rm SN}=0.4\times 10^{51}$~ergs). Blue and red lines correspond
to the models with four times lower ($0.1\times 10^{51}$~ergs) and
four times larger ($1.6\times 10^{51}$~ergs) values of the parameter
$E_{\rm SN}$, respectively.}
\end{figure}

\begin{figure}
\begin{center}\includegraphics[scale=0.5]{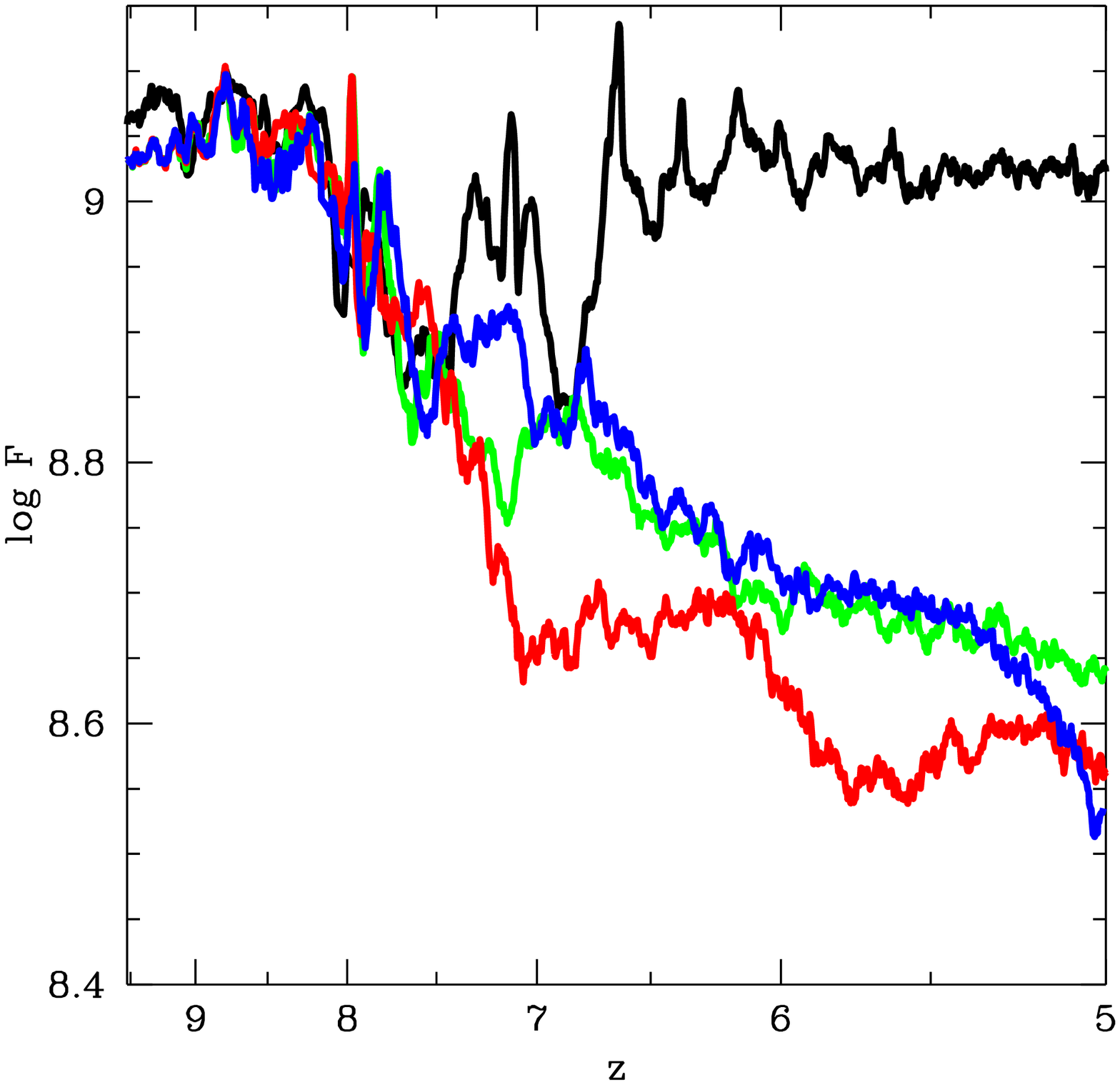}\end{center}
\caption{Evolution of the central dark matter phase space density, $F$, for 
small-box simulations with different values of $\kappa$. Black and
green lines correspond to the \dmo\ and hydrodynamic fiducial models
($\kappa=1$). Blue and red lines correspond to the models with
$\kappa=0.2$ and 5, respectively.}
\end{figure}

\begin{figure}
\begin{center}\includegraphics[scale=0.5]{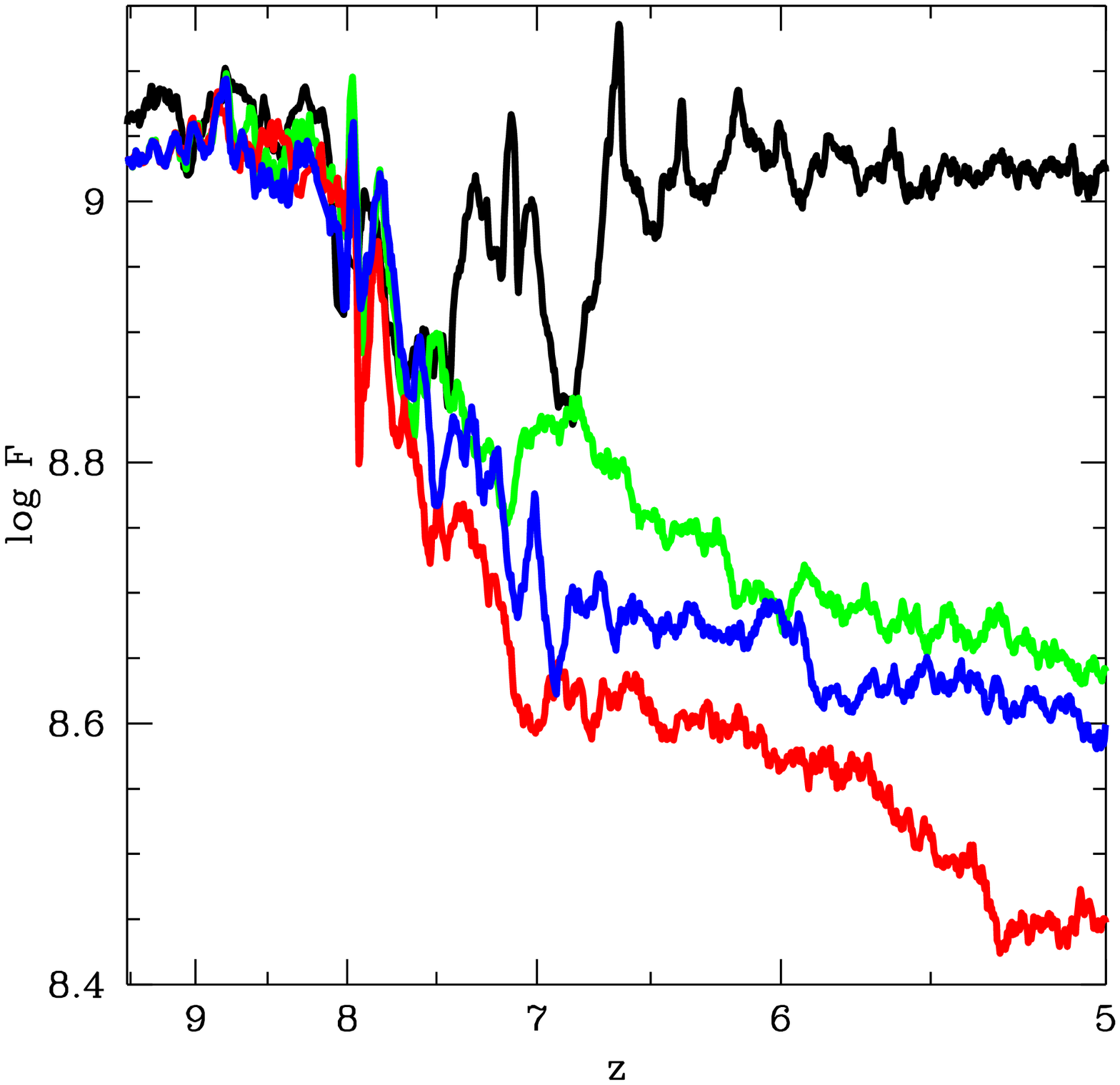}\end{center}
\caption{Evolution of the central dark matter phase space density, $F$, for 
small-box simulations with different values of $\xi$. Black and green
lines correspond to the \dmo\ and hydrodynamic fiducial models
($\xi=1$). Blue and red lines correspond to the models with $\xi=0.2$
and 5, respectively.}
\end{figure}

In Figures~S2--4 we show the evolution of the central dark matter
phase space density, $F$, measured in the same way as in the full-box
simulations (see the main text of the paper). Specifically, it was
measured for the central 5000 dark matter particles, typically located
within the central $\sim 200$~pc. This quantity is relatively
insensitive to the variable adiabatic compression of dark matter
caused by baryons (gas and stars). Black and green lines correspond to
fiducial values of the parameters for the \dmo\ and hydrodynamic
small-box simulations respectively (these lines are repeated in all
three figures).

A comparison of Figures~S2--4 with the bottom panel of Figure~2 from
the main text, shows: (a) for the \dmo\ models, in the small-box
simulations the quantity $F$ behaves similarly to the case of the
full-box simulations (converges to an approximately constant value);
and (b) gravitational dark matter heating due to stellar feedback
manifests itself in both large- and small-box simulations, but the
magnitude of the effect is smaller in the smaller box case. The latter
result is not unexpected, as in the small-box run the amount of dark
matter and gas available for the build-up of the model galaxy is much
smaller than in the full-box simulations, resulting in much less
energetic star formation after the initial strong star burst around
$z\simeq 7.7$.

Reducing the energy input from supernovae by a factor of four does not
affect appreciably gravitational heating of dark matter (see
Figure~S2). Four times larger energy input results in somewhat
stronger effect. The effect is clearly seen for the whole range of
$E_{\rm SN}$ tested.

\begin{figure}
\begin{center}\includegraphics[scale=0.5]{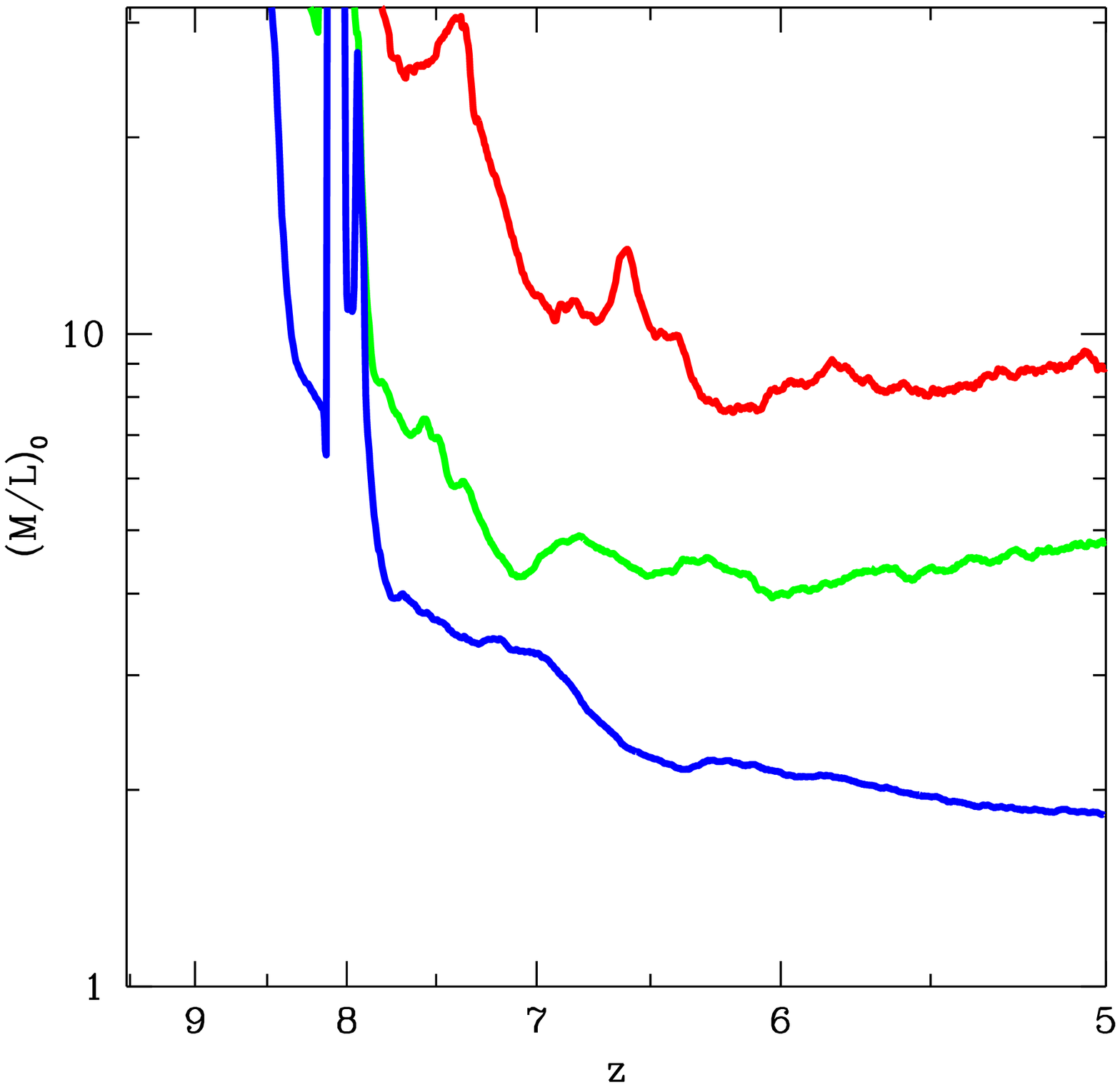}\end{center}
\caption{Evolution of the mass-to-light ratio within the central 200~pc, $(M/L)_0$, for
small-box simulations with different values of $E_{\rm SN}$. Blue,
green, and red lines correspond to the models with $E_{\rm
SN}=(0.1,0.4,1.6)\times 10^{51}$~ergs, respectively.}
\end{figure}

Figure~S5 demonstrates that we produce quite dark galaxies, with the
central value of the total mass-to-light ratio, $(M/L)_0$, reaching 2,
5, and 9 by $z=5$ for the models with $E_{\rm SN}=(0.1,0.4,1.6)\times
10^{51}$~ergs, respectively. (In calculating $(M/L)_0$ we adopted the
mass-to-light ratio of 1 for stars\cite{mat98s}, and ignored gas, as
current dwarf spheroidal galaxies have very little gas.) These values
are close to the observed ones; e.g., Leo~I and Fornax have
$(M/L)_0=3$ and 5, respectively\cite{mat98b}.

As discussed in \S~1, the low-temperature ($T<10^4$~K) radiative
cooling function we use in our model is based on several assumptions.
We investigated the sensitivity of the assumptions by running
small-box simulations with the original cooling function multiplied by
a constant factor, $\kappa$, equal to 0.2, 1, and 5.  The impact of
stellar feedback on the central phase space density of dark matter is
quite insensitive to the efficiency of low-temperature cooling as
shown in Figure~S4.  We emphasize, however, that to properly resolve
star formation and stellar feedback in dwarf galaxies, some form of
low-temperature cooling has to be present in the model. In our runs
with no radiative cooling for $T<10^4$~K, the shortest Jeans length
for gas particles inside the galaxy was $\sim 10$~kpc -- much larger
that the size of the galaxy. As a result, the galactic disk was
absolutely gravitationally stable, and any star formation prescription
produced stars at random locations inside the disk. In this case stars
are not formed in clusters and stellar feedback does not produce
large-scale gas bulk motions (instead heating the interstellar gas
almost uniformly). Not surprisingly, in such models the dark matter
experiences essentially no gravitational heating due to the feedback.

The last parameter we explored was the initial abundance of heavy
elements inside the model galaxy. The fiducial gas metallicity at
$z=9.6$ was $\sim 0.001$ solar units (see \S~1). We ran two small-box
models with the metallicity of each gas particle multiplied by a
factor $\xi$ equal to $0.2$ and 5. Figure~S5 shows that both the lower
and higher metallicity cases resulted in a somewhat stronger
gravitational heating effect. This may be due to either a significant
non-linear dependence of the strength of the effect on gas
metallicity, or (more likely) on the stochastic nature of our star
formation and stellar feedback prescriptions. Again, for the range of
$\xi$ tested, gravitational heating due to stellar feedback is
significant.

To summarize, the small-box simulations demonstrated that our main
result---that stellar feedback results in significant gravitational
heating of dark matter at the centers of primordial dwarf
galaxies---is largely insensitive to the values of the model
parameters $E_{\rm SN}$, $\kappa$, and $\xi$ when they vary within
reasonable ranges. We believe that this is because small galaxies are
effectively ``pressure cookers'' in which star formation and feedback
are confined to the central parts of the halos. A combination of model
parameters that results in more energy being pumped into the
interstellar gas will result in individual star bursts that have a
stronger effect on dark matter, but the interval between star bursts
becomes longer (it takes longer for hotter gas to cool down), so the
overall cumulative effect is qualitatively similar.

It follows from our numerical experiments that to accurately model
formation and evolution of a dwarf galaxy in a cosmological context,
the following conditions have to be met: (a) The model has to include
a prescription for low-temperature ($<10^4$~K) radiative gas cooling
due to heavy elements. This makes the simulations significantly (by at
least a factor of 10) more expensive, but is absolutely necessary if
one wants to to resolve the critical star formation and feedback
processes. (b) In SPH codes, the gas mass resolution has to be high
enough (a few hundred M$_\odot$, or better) to properly resolve the
extremely low density gas inside supernovae remnants: it is the
thermal pressure of this hot ($>10^6$~K) gas that drives the
large-scale bulk motions of the interstellar gas.

\end{document}